\documentclass[%
 aip,
 amsmath,amssymb,
 reprint,%
]{revtex4-1}

\usepackage{graphicx}
\usepackage{dcolumn}
\usepackage{bm}

\usepackage[utf8]{inputenc}
\usepackage[T1]{fontenc}
\usepackage{mathptmx}
\usepackage{etoolbox}

\newcommand{\VB}{$V_B^- \ $}
\newcommand{\dgs}{$D_{gs} \ $}
\newcommand{\egs}{$E_{gs} \ $}
\newcommand{\Bz}{$B_z \ $}
\newcommand{\nuplus}{$\nu_2 \ $}
\newcommand{\numinus}{$\nu_1 \ $}
\newcommand{\etadc}{$\eta_{DC} \ $}

\newcommand{\deltaBmax}{$\Delta B_{z}^{max} \ $}
\newcommand{\micra}{$\mu \text{m} \ $}
\newcommand{\micrasqr}{$\mu \text{m}^2 \ $}

\makeatletter
\def\@email#1#2{%
 \endgroup
 \patchcmd{\titleblock@produce}
  {\frontmatter@RRAPformat}
  {\frontmatter@RRAPformat{\produce@RRAP{*#1\href{mailto:#2}{#2}}}\frontmatter@RRAPformat}
  {}{}
}%
\makeatother
\begin{document}

\preprint{AIP/123-QED}

\title[Real-time ESR tracking for sub-micron 3D magnetic mapping with \VB quantum sensors in hexagonal boron nitride]{Real-time ESR tracking for sub-micron 3D magnetic mapping with \VB quantum sensors in hexagonal boron nitride}
\author{Jefferson A. O. Galindo*}\email{jefferson.galindo@ufpe.br}

\affiliation{Department of Physics, Universidade Federal de Pernambuco, Recife, 50740-540, Pernambuco, Brazil}    

\author{Edwin D. C. Sanchez}%
\affiliation{Department of Physics, Universidade Federal de Pernambuco, Recife, 50740-540, Pernambuco, Brazil}

\author{Cecília L. A. V. Campos }
\affiliation{Department of Physics, Universidade Federal de Pernambuco, Recife, 50740-540, Pernambuco, Brazil}
\author{Allison R. Pessoa}
\affiliation{Department of Physics, Universidade Federal de Pernambuco, Recife, 50740-540, Pernambuco, Brazil}
\affiliation{Instituto Federal de Educação, Ciência e Tecnologia de Pernambuco, Recife, 50740-540, Pernambuco, Brazil}

\author{Hugo A. D. Correia}
\affiliation{Física de Materiais, Escola Politécnica de Pernambuco, Universidade de Pernambuco, Recife, 50720-001, Pernambuco, Brazil.}

\author{José D. M. de Lima}
\affiliation{Department of Physics, Universidade Federal de Pernambuco, Recife, 50740-540, Pernambuco, Brazil}

\author{Klaus Krambrock}
\affiliation{Departamento de Física, Universidade Federal de Minas Gerais, Belo Horizonte, 31270-901, Minas Gerais, Brazil}

\author{Leonardo de S. Menezes}
\affiliation{Chair in Hybrid Nanosystems, Faculty of Physics, Ludwig-Maximilians-Universität München, 80539, Bavaria, Germany}
\affiliation{Department of Physics, Universidade Federal de Pernambuco, Recife, 50740-540, Pernambuco, Brazil}

\author{Anderson M. Amaral}
\affiliation{Department of Physics, Universidade Federal de Pernambuco, Recife, 50740-540, Pernambuco, Brazil}

\date{\today}

\begin{abstract}
The discovery of spin-dependent luminescent properties of negatively charged boron-vacancy centers ($V^-_B$) in hexagonal boron nitride (hBN) enabled a new platform for quantum sensing with van der Waals materials. Particularly, the possibility of performing optically detected magnetic resonance (ODMR) for determining the electron spin resonance (ESR) frequencies of hBN color centers became a strong tool for quantum sensing of magnetic fields with submicrometric resolution. However, due to low ODMR contrast, current techniques proposed for mapping DC magnetic fields require hours of integration to obtain a magnetic image of a micron-sized region. In this work, we report the implementation of a frequency-tracking approach for real-time monitoring of ESR frequencies of localized $V^-_B$ centers in hBN. With this technique, magnetic field monitoring was used to map the field pattern generated by a micron-sized conical magnetic tip in only a few minutes. By controlling the magnetic sample's position relative to the quantum sensor, three-dimensional magnetic mapping of the field was achieved with diffraction-limited resolution and shot-noise-limited sensitivity of 54 $\mu$T$/\sqrt{\text{Hz}}$. Magnetic field gradients of 3.6 $\pm$ 0.2 $\mu$T/nm were measured with our system, in which a maximum detected field rate of 6 mT/s was achieved. The results of this study establish spin resonance frequency-tracking as a viable technique and fast method for minute-scale magnetic imaging, reducing acquisition times by at least one order of magnitude if compared to conventional techniques.
\end{abstract}

\maketitle




The growing demand for quantum sensors utilizing solid-state optically active materials has driven the development of a variety of sensing platforms at the sub-micron scale \cite{roberts2025quantum,awschalom2018quantum}. Among these systems, color centers in diamond and silicon carbide have been extensively explored \cite{jelezko2006single,rose2018observation,widmann2015coherent,wu2025advances}, with negatively charged nitrogen-vacancy color centers in diamond being the most investigated system for sensing applications to date \cite{paudel2024,kenny2025,villing2025}.

However, recent developments on luminescent color centers in bidimensional (2D) van der Waals (vdW) materials have gained attention due to their promising capabilities as quantum sensors \cite{fang2024quantum,jana2026two,melendez2025quantum}. 
Advantages of using vdW materials rely first on their inherent atomic-scale thickness, which ensures high spatial resolution, inherent proximity to the target sample, and strong coupling to near fields \cite{fang2024quantum}. Second, the layered nature of the sensor mitigates coherence degradation due to surface inhomogeneities, charge fluctuations, and spin noise, as already reported in nanodiamonds \cite{wang2026_atomic_thin,romach2015,sangtawesin2019}. 

A prominent example of a vdW material is hexagonal boron nitride (hBN), which hosts color centers with optically addressable spin suitable for quantum sensing\cite{gottscholl2021spin,wang2026_atomic_thin}. In hBN, negatively-charged boron vacancy centers\cite{gottscholl2020initialization} ($V_B^-$) and carbon-related centers \cite{mendelson2021carbon} have already been used as sensors for magnetic fields  \cite{vaidya2023quantum,gottscholl2020initialization,gottscholl2021spin,choi2024FET}, temperature and pressure \cite{muhammad2024anisotropic}, gases \cite{goel2021gassensor,das2025}, spin wave excitations \cite{Zhou2024spin}, strain \cite{lyu2022strain}, and neutron detection \cite{somasundaram2025_fast}. Recent works on magnetic sensing with \VB \cite{kumar2022,huang2022wide,healey2023quantum,sasaki2023imaging,ma2025imaging,mu2025magnetic} reported magnetic field imaging based on optically detected magnetic resonance (ODMR) techniques, resulting in diffraction-limited spatial resolution maps with sensitivities up to 60 $\mu$T$/\sqrt{\text{Hz}}$ at room temperature \cite{kumar2022}. However, the techniques commonly used for magnetic field imaging are based on the acquisition of a complete ODMR spectrum at each pixel and on wide-field imaging with CCD cameras. In both methodologies, the sample is static relative to the quantum sensor, and long integration times ($>$ 10 h) are necessary  \cite{huang2022wide,mu2025magnetic}.  

Here, we employ a frequency-tracking method for magnetic imaging with \VB color centers that records a complete map in only a few minutes. Our approach is based on real-time electron spin resonance (ESR) frequency tracking under microwave modulation, which enables continuous, real-time monitoring of the local magnetic field. As a demonstration, we present the first three-dimensional (3D) magnetic mapping of the remnant field generated by a sharp metallic tip with a micron-sized apex, obtained through 3D spatial control of the magnetic sample.

\begin{figure}
    \centering
    \includegraphics[width = 1.0\linewidth]{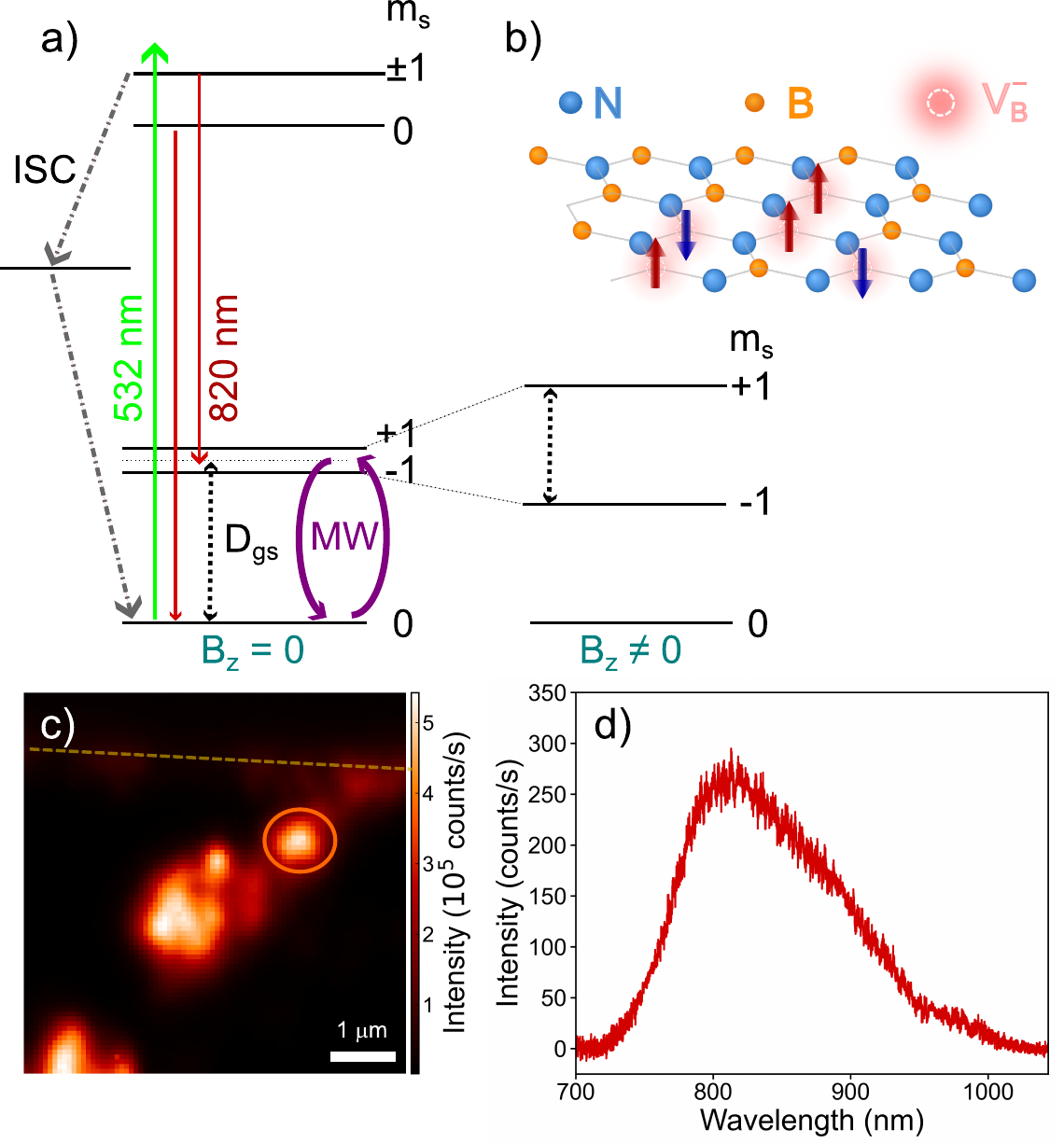}
    \caption{a) Simplified energy level diagram for the \VB color centers. The \VB centers are excited with 532 nm laser light (green arrow) and emit a broad luminescence band centered at 820 nm (red arrows). Nonradiative decay (dashed-dotted arrows) from the $m_s = \pm1$ states via intersystem crossing through the metastable state leads to spin-dependent emission intensity. Ground-state zero-field splitting \dgs is represented. In the presence of a nonzero magnetic field $B_z$, the Zeeman splitting sums over $E_{gs}$. A MW signal around 3.45 GHz (purple arrows) is used to coherently manipulate the \VB ground-state spin. b) Atomic arrangement of hBN containing \VB color centers. c) Luminescence 6$\times$6 $\mu$m$^2$ confocal raster-scan of the sample where the selected \VB ensemble (orange circle) and the gold antenna's edge (dashed yellow line) can be noticed. d) \VB emission spectra under 532 nm excitation.}
    \label{fig:Fig1}
\end{figure}

In our experiments, localized ensembles of \VB in bidimensional hBN\cite{gottscholl2020initialization,gottscholl2021spin} were the chosen quantum sensors. 
The \VB color center consists of a missing atom of boron associated with three equivalent nitrogen atoms in the hBN matrix. At room temperature, \VB presents $D_{3h}$ point-group symmetry in which both ground and excited states are spin triplets [Fig. \ref{fig:Fig1}a)]. The atomic arrangement of the \VB color centers in hBN is shown in Fig. \ref{fig:Fig1}b). In the ground state, the color center presents a zero-field splitting (ZFS) \dgs $\sim3.45$ GHz, with a transverse ZFS \egs $\sim 100$ MHz, that naturally lifts the degeneracy of the $m_s = \pm 1$ spin states \cite{gottscholl2021spin}. In the presence of a nonzero external magnetic field ($B_z$) aligned to the spin quantization $c \parallel z$ axis of hBN, the $m_s = + 1$ and $m_s = - 1$ states are subjected to Zeeman splitting. 
Due to the coupling of the $m_s = \pm 1$ spin states of the excited state to a metastable state via intersystem crossing, the color center's luminescence around 820 nm is spin-dependent under linearly polarized 532 nm laser light excitation\cite{gottscholl2020initialization,gottscholl2021spin}. This photophysical feature allows the implementation of the ODMR technique to determine the spin projection of the color center via coherent manipulation carried out by an external microwave (MW) electromagnetic field\cite{gottscholl2021spin}. 

The samples investigated consisted of neutron-irradiated hBN flakes fabricated from ultrafine hBN powder ($ \sim 70$ nm), prepared in a nuclear reactor with a thermal flux of $4\times 10^{12}$ neutrons cm$^{-2}$ s$^{-1}$ during 16 h \cite{krambrock2018_samples}. 
A home-made confocal microscope equipped with a high-numerical-aperture objective (N.A. = 1.4) and an avalanche photodiode (APD) connected to a photon counting device was used to detect the luminescent \VB ensembles. The hBN flakes are deposited by spin coating on a glass coverslip containing a photolithographed gold MW antenna for spin manipulation (Fig. S1). 
Fig. \ref{fig:Fig1}c) presents a raster scan of the sample, in which localized \VB ensembles can be identified. 
By controlling the sample's position, we selected a given set of color centers close to the antenna's edge for further measurements. The emission spectrum of the \VB is shown in Fig. \ref{fig:Fig1}d). 
The \VB ensembles exhibit a broad emission band centered at 820 nm, which is spectrally filtered from the excitation by using a set of longpass filters.
In addition, strong luminescence from \VB was detected, with detected photon rates up to $\sim 8 \times 10^5$ counts/s under 530 $\mu$W excitation power with no detected optical power broadening (Fig. S2).

\begin{figure*}
    \centering
    \includegraphics[width=.7\linewidth]{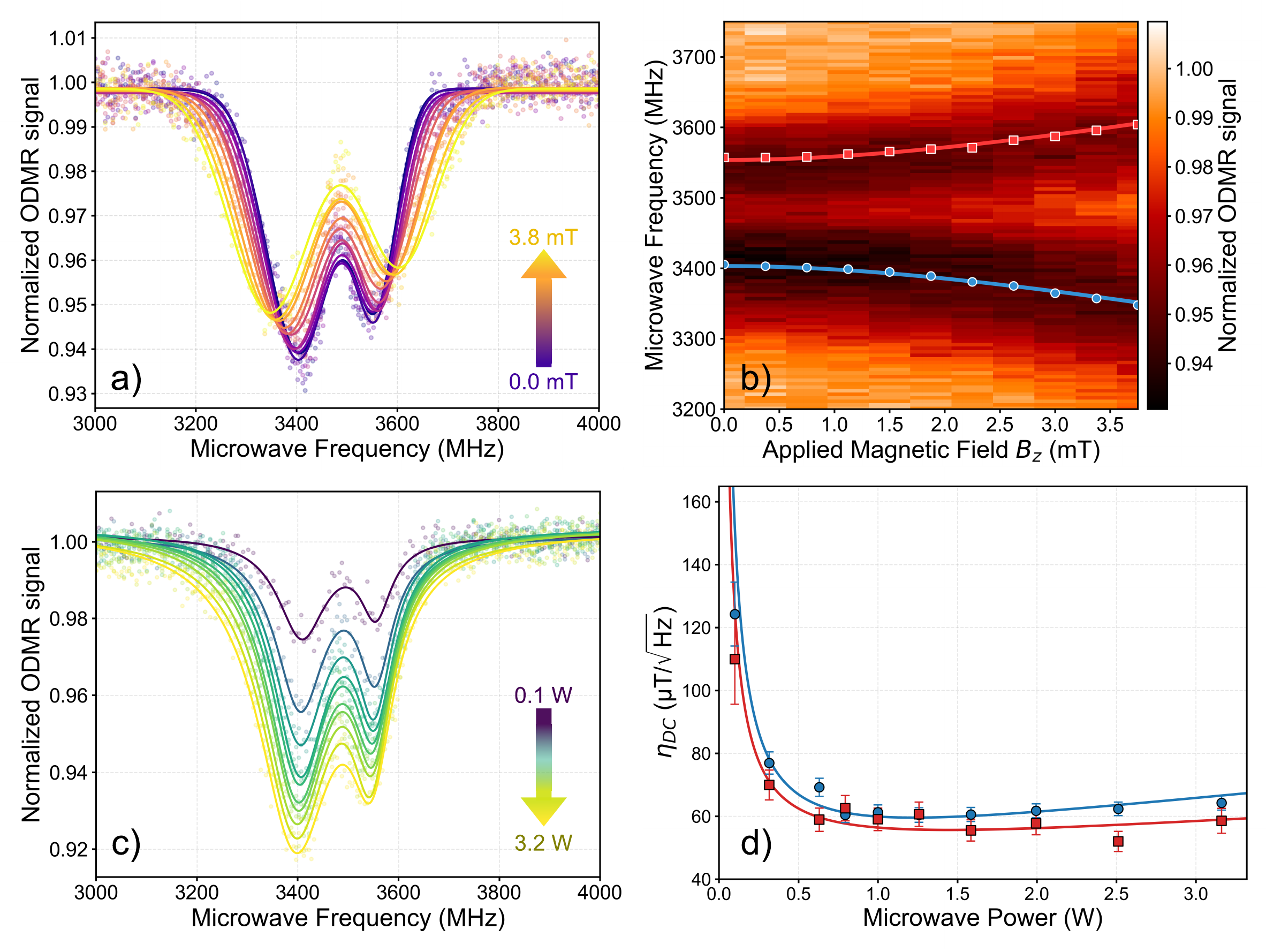}
    \caption{a) ODMR spectra as a function of magnetic field $B_z$ measured under 1 W of MW power. The data (dots) were fitted with Gaussian curves (solid lines).
    b) ODMR colormap representation with the same data points presented in a).  Bright areas (dark areas) represent high (low) ODMR signal. The measured \numinus (blue dots) and \nuplus (red squares) are shown along with respective fits (solid lines) using Eq. (\ref{eq:nupm}).
    c) ODMR spectra with increasing MW power taken with \Bz$=0$. 
    d) Magnetic field DC sensitivity \etadc using the \numinus (blue dots) and \nuplus (red squares) ESRs as a function of MW power, with blue and red solid lines being the respective fits obtained using Eq. (\ref{eq:etadc}) at zero external magnetic field. For both ESRs, \etadc reaches around 60 $\  \mu$T/$\sqrt{\text{Hz}}$ above 0.7 W.}
    \label{fig:Fig2}
\end{figure*}

The ODMR spectra of the selected ensemble as a function of the external magnetic field parallel to the hBN $c$ axis are presented in Fig. \ref{fig:Fig2}a). For this purpose, ODMR measurements were carried out as a function of the external field \Bz generated by a single coil positioned over the sample at a distance of 0.5 mm from the glass substrate. The data were fitted with two Gaussian curves, and all measurements were taken at $1$ W of MW power. 
For better visualization of the Zeeman-induced shifts on the ESR center frequencies due to $B_z$, a colormap representation of the ODMR curves is shown in Fig. \ref{fig:Fig2}b). Each vertical line represents the ODMR signal at a given $B_z$, with darker areas indicating ODMR dips centered at ESR frequencies. Blue dots and red squares are the measured ESR frequencies \numinus and \nuplus that correspond to transitions to the $m_s = -1$ and $m_s = +1$ states, respectively. At each measurement, the full width at half maximum (FWHM) and contrast $C$ of each ESR were also obtained (Table S1). 

At \Bz$= 0$, the selected ensemble presented a $C = $ 6.4$\%$, typical for \VB color centers. When $B_z$ increases, the ESRs shift apart due to Zeeman splitting, and their corresponding curves become broadened [Figs. \ref{fig:Fig2}a) and b)]. 
In contrast, no frequency shift of the ESRs was observed under applied in-plane fields ($B_x$ and $B_y$) up to 20 mT (Fig. S3)
This effect may be related to the inhomogeneous broadening of the \VB spins caused by slight deviations from the $c$ axis of distinct sites or layers of the hBN flake and strain \cite{murzakhanov2021_B_broadening}. 

The relation between the ESRs and the $B_z$ field for the \VB ground state can be obtained from the standard spin Hamiltonian for spin-1 systems\cite{gottscholl2020initialization}

\begin{equation}
    H = h \left\{ D_{gs} [ S_z^2 - S(S+1)/3] +  E_{gs}(S_x^2-S_y^2)+ \gamma_e\vec{B} \cdot \vec{S} \right\},
\label{eq:hamilt}
\end{equation}

\noindent where $h$ is the Planck's constant, $\vec{S}$ is the total electron spin for the triplet state ($S=1$), $S_{x,y,z}$ are the spin-1 operators, $\gamma_e$ is the gyromagnetic ratio for the electron equal to 28 MHz/mT for $V_b^-$, and $\vec{B}$ is the applied magnetic field. For a static magnetic field applied parallel to the $c$ symmetry axis of hBN, \numinus and \nuplus can be obtained from Eq. (\ref{eq:hamilt}) and written as
\begin{equation}
    \nu_{1,2} = D_{gs} \mp \sqrt{E_{gs}^2 +  (\gamma_e B_z)^2}.
\label{eq:nupm}
\end{equation}
From the ODMR spectra data with increasing $B_z$, the increasing separation between \numinus and \nuplus becomes evident [Fig. \ref{fig:Fig2}b)]. An excellent fit can be obtained by using Eq. (\ref{eq:nupm}), with $g=2.000$, $D_{gs} = 3.48$ GHz, and $E_{gs} = 75$ MHz, confirming previous results reported for \VB centers \cite{gottscholl2020initialization,gottscholl2021spin}.

The ODMR signal as a function of the MW power dissipated at the antenna is shown in Fig. \ref{fig:Fig2}c). The ODMR contrast for the \numinus ($\nu_2$) ESR increases from 2.6$\%$ (1.9$\%$) to a saturated contrast of 7.9$\%$ (4.3$\%$). Additionally, power broadening \cite{hussain_2026_power_broadening,haykal2022decoherence,zhou2023_power_broadening} of the ESR lines with increasing MW power is noticeable (Table S2 and Fig. S4). 

In magnetic sensing, it is important to discuss the DC field sensitivity \etadc of \VB centers under investigation \cite{gottscholl2021spin}. In this sense, \etadc is the figure of merit used for characterizing ODMR-based quantum sensors, written as
\begin{equation}
    \eta_{DC} = \mathcal{P} \times \frac{1}{\gamma_e} \times \frac{\Delta \nu}{C\sqrt{R}},
    \label{eq:etadc}
\end{equation}
in which $\mathcal{P} = 0.70$ is a parameter related to the ODMR lineshape profile (0.7 for a Gaussian lineshape), $\Delta\nu$ is the Gaussian FWHM for a given ESR line, and $R$ is the rate of detected photons.

As mentioned above, by increasing \Bz and MW power, the ESR lineshapes are broadened, which affects $\eta_{DC}$. During a real-time measurement of $B_z$, one has no control over how the field changes during the experiment. Thus, the optimization of \etadc with MW powers must be carried out to optimize the system's sensing capabilities. For this purpose, we measured \etadc as a function of MW power [Fig. \ref{fig:Fig2}d)]. 
As power increases, the FWHMs also increase, but the contrast is enhanced, reducing the denominator value in Eq. (\ref{eq:etadc}) before $C$ saturates. Consequently, \etadc goes from $(110 ~\pm~ 10)$ $\mu$T/$\sqrt{\text{Hz}}$ and $(120 ~\pm~ 11)$ $\mu$T/$\sqrt{\text{Hz}}$ to an approximately constant value of $(63 ~\pm ~ 3) \ \mu$T/$\sqrt{\text{Hz}}$ and $(59 ~\pm ~ 4) \ \mu$T/$\sqrt{\text{Hz}}$ above $0.7$ W for \numinus and \nuplus, respectively. 
For field sensing purposes, due to the low $C$ of the ODMR signal, if compared to nitrogen-vacancy centers ($C\sim 20\%$) \cite{rondin2012tracking}, a tradeoff between MW power and $C$ must be struck to reduce integration times and to avoid accentuated heat dissipation at the antenna, which can shift the ESR frequencies by $dD_{gs}/dT = -0.9$ MHz/K at room temperature \cite{liu_temperature2025}. In our experiments, an average shift of $8$ MHz was observed between the measured ESR centers at $0.1$ W and $3.2$ W, indicating a temperature increase of 9 K due to heat dissipation. However, once thermal equilibrium was reached at a fixed MW power level, no significant frequency drift was observed. Since all magnetic scans were performed after thermal stabilization and at constant optical and MW powers, shifts in the tracked \nuplus were attributed solely to variations in $B_z$.


\begin{figure*}
    \centering
    \includegraphics[width=\textwidth]{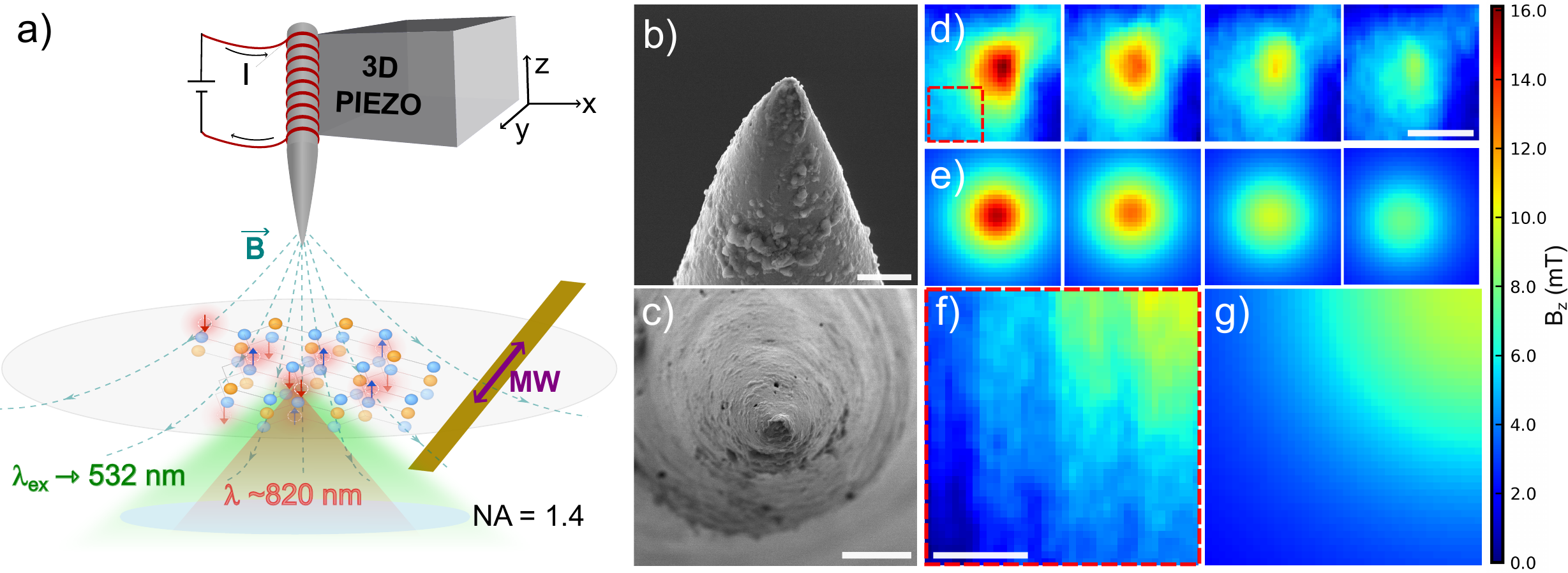}
    \caption{a) Representation of the 3D magnetic field mapping experimental setup. Laser light at 532 nm is used to excite the quantum sensor (\VB ensemble), and part of the emitted luminescence (820 nm) is collected by an objective lens (N.A. = 1.4) and sent to the detection system. 
    During magnetic mapping, the sensor remains stationary while the magnetic tip is scanned in the $xy$ plane using a 3D piezoelectric stage at a fixed distance to sample the $z$ direction. By setting the current intensity (I) carried by a coil surrounding the tip's body, the magnetization is controlled.
    b) Side view and c) bottom view scanning electron microscopy images of the conical magnetic tip with respective scale bars of 10 $\mu$m and 20 $\mu$m. 
    d) 3D Magnetic field mapping of the sample constructed from 20$\times$20 \micrasqr maps in the $xy$ plane at distinct $z$ positions. The images were taken in 400 nm steps, with a 2D magnetic scan taken at each new $z$ position. The scale bar represents 10 $\mu$m. e) Simulated magnetic maps obtained from measured tip parameters. f) measured and g) simulated 6$\times$6 $\mu$m$^2$ scan at the closest tip-to-sample position (red dashed box in d). The scale bar corresponds to 2 $\mu$m. }
    \label{fig:Fig3}
\end{figure*}

For real-time magnetic field monitoring with \VB color centers (Fig. S5), we have adapted the widely investigated ESR frequency-tracking technique for sensing with nitrogen-vacancy color centers in diamond\cite{rondin2012tracking,maletinsky2012robust,tettienne2014}, and previously reported by our group \cite{edwin2020}. With this technique, while tracking \nuplus, a best shot-noise-limited sensitivity\cite{barry_2024_shot} of 54 $\mu$T$/\sqrt{\text{Hz}}$ (Fig. S6) was achieved for an integration time of 20 ms and 1 W of MW power, in agreement with the measured \etadc [Fig. \ref{fig:Fig2}d)]. We were also able to track slow sinusoidal \Bz fields with frequencies up to 0.4 Hz (Fig. S7). 

The scheme of the sensing protocol used in the 3D magnetic field map measurements of a micron-sized ferromagnetic conical tip is presented in Fig. \ref{fig:Fig3}. In our setup, the luminescent \VB ensemble with an optical detection area of 0.06 $\mu\text{m}^2$ is our static quantum sensor. 
A micron-sized tip (apex of 2.4 $\mu$m diameter) was fabricated from a commercial steel wire by electrochemical etching with an HCl solution\cite{zhang_2021_etching} [Figs. \ref{fig:Fig3}b) and c)]. The tip is positioned near the hBN flake using a moving mount equipped with a micrometric positioner and a 3D piezoelectric stage. 
The tip's magnetization is controlled by the current flowing through a wire coil surrounding the cylindrical tip body. By characterizing the magnetic hysteresis of the steel wire and by knowing the magnitude of the magnetic field induced by the coil (Fig. S8), we calculated that a magnetization of 75 kA/m is generated along the steel wire for a current of 1.0 A. 

Once the tip is manually positioned over the sample, the frequency-tracking system is initialized for monitoring \nuplus until a maximum is reached, corresponding to a maximum detected $B_z$. Then, we slowly approach the tip using the piezoelectric stage until contact with the sample's surface is detected. At this point, the tip is moved away 1 \micra from the surface to a safe working distance. The contact does not affect the tip's shape, as confirmed by the SEM images [Figs. \ref{fig:Fig3}b) and c)] taken after the measurements.

For 3D magnetic field mapping, the tip was scanned in the $xy$-plane to acquire 20$\times$20 \micrasqr magnetic maps, as depicted in Fig. \ref{fig:Fig3}d). The tip started at $\sim$ 1 \micra and was moved away from the sensor in 400 nm steps along the $z$-axis, and a new map was recorded at each position. With an integration time per pixel of 400 ms and a grid of 30$\times$30 pixels, each map required approximately 6 minutes of acquisition.
%
To filter spurious noise, a Savitzky-Golay filter of order 2 was applied to the final image. The $B$ color bar stands for all maps and indicates a maximum measured \deltaBmax of (16.4 $\pm$ 0.5) mT at the closest tip-to-surface arrangement. 

From the measured maps, one can notice the expected circular patterns of the tip, with increasing intensity at the center. Inhomogeneities in the circular magnetic patterns can be associated with asymmetric corrosion of the tip and non-symmetric remains close to the tip's apex, as confirmed by SEM images [see Fig. \ref{fig:Fig3}b)] and measurements for a distinct tip (Fig. S9).

For comparison with the obtained data, we fitted the measured magnetic fields with those generated by an ideal conical tip with the same geometric dimensions.  Only three parameters were fitted from data in the maps: the tip center's position, the tip-to-surface distance ($z$), and the overall magnetization of the tip ($M$). The results are presented in Fig. \ref{fig:Fig3}e).  A good fit of the overall intensities \Bz and the central positions of the tip was obtained. From the fit, the average $M$ of (90 $\pm$ 32) kA/m was consistent with the magnetization of the sample (see section S7). The fitted $z$ positions ranged from (0.5 $\pm \ 0.8$) \micra to (3.0 $\pm \  0.8$) \micra, in agreement with the experiment. Large errors are attributed to the simplicity of our monopole model, which does not perfectly capture the irregular geometry of the tip apex observed in the SEM images [Figs. \ref{fig:Fig3}b) and c)].

To explore the resolution limits of our system, a small scan [Fig. \ref{fig:Fig3}f)] was carried out at minimum $z$ in the box depicted in Fig. \ref{fig:Fig3}d).  In this scan, the pixel size is 200 nm (30$\times$30 pixels), close to the 287 nm diffraction limit of our system \cite{edwin2020}. The results correspond well to the simulated magnetic field shown in Fig. \ref{fig:Fig3}g), where the circular gradient matches the measured data.  
From the results shown in Figs. \ref{fig:Fig3}d) and f), we obtained a maximum field gradient \deltaBmax of (3.6 $\pm$ 0.2) $\mu$T/nm. Our system was able to track magnetic field rates of (6.0 $\pm$ 0.4) mT/s. These results indicate that the quantum sensors are suitable for measuring field gradients of a few $\mu$T/nm per second with submicrometric precision. 

Furthermore, from the magnetic field maps, the dependence of \Bz on the tip's distance to the sample was investigated. By plotting the maximum measured field $B_z^{max}$ of each image as a function of $z$, we could obtain the inverse quadratic dependence of the field with distance for a point-like magnetic monopole generated by conical tips \cite{rugar1990magnetic} (Fig. S10). The sensitivity and resolution reached by our system are comparable to those found in the literature \cite{kumar2022,huang2022wide,healey2023quantum,sasaki2023imaging,mu2025magnetic}, with the proposed technique leading to considerably shorter integration times and spatial control of magnetic sample position relative to the quantum sensor. It is worth noticing that the sensing parameters obtained in this study can be improved by implementing single \VB color centers\cite{gilardoni2025_nat_com,bhattacharya2026}, dynamical decoupling techniques\cite{rizzato2023extending}, AFM-assisted quantum sensor nanoprobe\cite{maletinsky2012robust},  and optical and MW optimization for ODMR sensing\cite{ren2025optimization}.

In summary, we reported a detailed study on the characterization of \VB color centers in 2D hBN as quantum sensors for DC magnetic field monitoring. Despite the low ODMR contrast compared to other color centers\cite{jelezko2006single}, the optical robustness of the \VB and the high photon emission rates allowed us to implement a frequency-tracking technique for real-time magnetic field monitoring.
The results show that our system enables fast magnetic imaging with gradients of a few $\mu$T/nm generated by a micrometric structure. Unlike conventional methods that require hours to acquire a single map, our approach reduces this time to a few minutes. The spatial control of the magnetic sample relative to the sensor allowed us to reach diffraction-limited submicron resolution and a best shot-noise-limited sensitivity of 54 $\mu$T$/ \sqrt{\text{Hz}}$. Furthermore, consecutive magnetic imaging at distinct sample positions highlights the capabilities of \VB color centers for constructing 3D magnetic field maps. Applications of the proposed method can be extended to magnetic mapping of micro-circuits \cite{garsi2024three} and magnetic nanoparticles \cite{sadzak2018coupling,mathes2024nitrogen}, hybrid sensing \cite{wang2018_hybrid} and nm resolution sensing with single color centers \cite{gilardoni2025_nat_com}. 

We invite the distinguished readers to take a look at the supplementary material for more information on the experimental setup, excitation power dependence on luminescence and ESR linewidth. The effects of in-plane oriented external magnetic fields on ODMR, details on the frequency tracking method, and shot-noise-limited sensitivity are also described there. Finally, the description of the simulated magnetic fields,  field maps of additional tips, and the dependence of the measured $B_z$ on $z$ are presented.

This research was supported by the program MCTI/FINEP/FNDCT/-Centros Temáticos 2023 (grant number 1020/24), Petrobras, INCT-INFo for financial support, CeNS, Ludwig-Maximilians-Universität München, and the Bavarian program EQAP.
\section*{Author Declarations}
\subsection*{Conflicts of interest}
The authors have no conflicts of interest to disclose.
\subsection*{Author Contributions}
\textbf{Jefferson A. O. Galindo:} Conceptualization (equal);
Data curation (lead); Formal analysis (lead); Investigation (equal); Methodology (equal); Software (lead); Writing – original draft (lead); Writing – review \& editing (equal).
\textbf{Edwin D. C. Sanchez:} Conceptualization (equal); Data curation (equal); Formal analysis (equal); Investigation (equal); Methodology (equal); Validation (equal); Resources (equal); Writing – original draft (equal); Writing – review \& editing (equal).
\textbf{Cecília L. A. V. Campos:} Conceptualization (equal); Investigation (equal); Resources (equal); Writing – review \& editing (equal).
\textbf{Allison R. Pessoa:} Conceptualization (equal); Investigation (equal); Methodology (equal); Validation (equal); Writing – original draft (supporting); Writing – review \& editing (equal).
\textbf{Hugo A. D. Correia:}  Investigation (equal); Methodology (equal); Validation (equal).
\textbf{José D. M. de Lima:} Investigation (equal); Methodology (equal); Resources (supporting).
\textbf{Klaus Krambrock:} Resources (equal); Validation (equal); Writing – review \& editing (equal).
\textbf{Leonardo de S. Menezes:} Project administration (equal); Supervision (equal); Funding acquisition (equal); Writing – review \& editing (equal).
\textbf{Anderson M. Amaral:} Conceptualization (equal); Investigation (equal); Project administration (equal); Supervision (equal); Funding acquisition (equal); Visualization (equal); Writing – original draft (supporting); Writing – review \& editing (equal).

\section*{Data Availability Statement}

The data resulting from the experiments and simulations carried out in this study are available from the corresponding author upon reasonable request. 

\section*{References}
\bibliography{refs}

\clearpage
\onecolumngrid
\pagestyle{empty}
 
\newcommand{\suppfile}{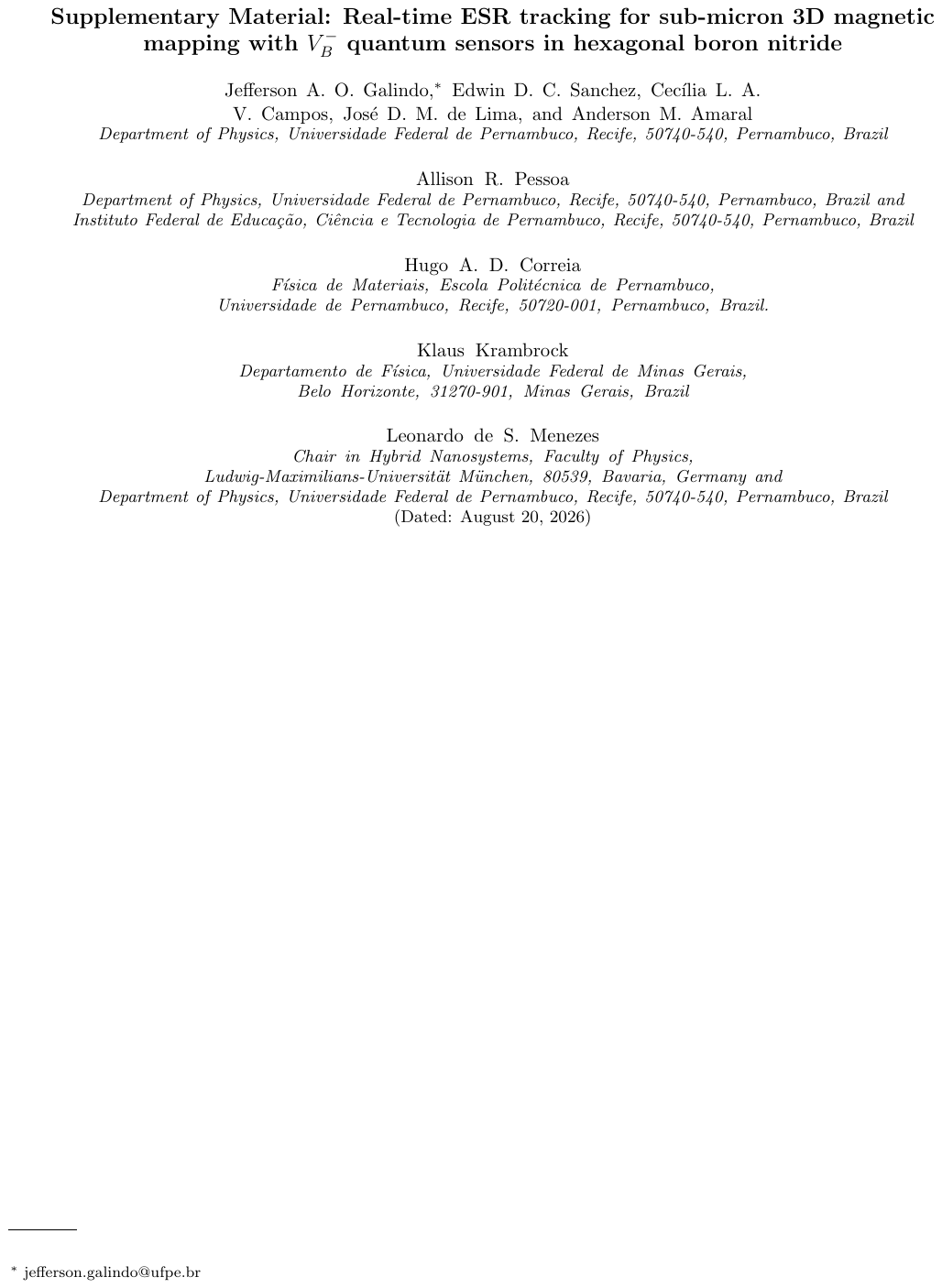}  
\newcount\supptotal \supptotal=12           
 
\newcount\suppn \suppn=1
\loop\ifnum\suppn<\numexpr\supptotal+1\relax
  \clearpage
  \thispagestyle{empty}%
  \noindent\hspace*{\dimexpr-1in-\oddsidemargin\relax}%
  \vspace*{\dimexpr-1in-\topmargin-\headheight-\headsep\relax}%
  \includegraphics[page=\the\suppn,width=\paperwidth]{\suppfile}%
  \advance\suppn by 1
\repeat

\end{document}